\documentclass[pra,aps,reprint,superscriptaddress,showpacs]{revtex4-1}

\usepackage{graphicx}
\usepackage{verbatim}
\usepackage{comment}
\usepackage{amsmath}
\usepackage{amssymb}
\usepackage{color}
\usepackage{xfrac}
\usepackage{units}

\newcommand{\eq}[1]{Eq.~\eqref{#1}}

\newcommand{\eqs}[1]{Eqs.~\eqref{#1}}

\newcommand{\ph}{\varphi}
\newcommand{\w}{\omega}
\newcommand{\ad}{\hat{a}^\dagger}

\DeclareMathOperator{\Tr}{Tr}

\begin{document}

\title{Optimization of finite-size errors in finite-temperature calculations \\ 
of unordered phases}

\author{Deepak Iyer} \affiliation{Department of Physics, The
  Pennsylvania State University, University Park, Pennsylvania 16802,
  USA} \author{Mark Srednicki} \affiliation{Department of Physics,
  University of California, Santa Barbara, California 93106, USA}
\author{Marcos Rigol} \affiliation{Department of Physics, The
  Pennsylvania State University, University Park, Pennsylvania 16802,
  USA}

\begin{abstract}
  It is common knowledge that the microcanonical, canonical, and grand
  canonical ensembles are equivalent in thermodynamically large
  systems. Here, we study finite-size effects in the latter two
  ensembles.  We show that contrary to naive expectations, finite-size
  errors are exponentially small in grand canonical ensemble
  calculations of translationally invariant systems in unordered 
  phases at finite temperature. Open boundary conditions and
  canonical ensemble calculations suffer from finite-size errors that
  are only polynomially small in the system size. We further show
  that finite-size effects are generally smallest in numerical linked
  cluster expansions.  Our conclusions are supported by analytical and
  numerical analyses of classical and quantum systems.
\end{abstract}

\maketitle

\section{Introduction}
\label{sec:introduction}

The use of identically prepared systems, or ensembles, has been
essential to our understanding of equilibrium and far-from-equilibrium
properties of classical and quantum systems. Traditionally, three
types of ensembles are used---(a) the microcanonical ensemble, which
involves systems with fixed energy and particle number, (b) the
canonical ensemble (CE), which involves systems with fixed particle number
in contact with a large reservoir (at temperature $T$) with which they
can exchange energy, and (c) the grand canonical ensemble (GE), which
involves systems in contact with a reservoir with which they can
exchange energy and particles (in equilibrium, the average particle
number is determined by the chemical potential $\mu$). Whereas these
three ensembles pose fundamentally different physical constraints, 
it can be shown that they are equivalent in the thermodynamic limit
(provided, of course, that temperatures and chemical potentials are
selected appropriately). Being technically easier to deal with, the
canonical and grand canonical ensembles are the most commonly used
ensembles in the literature. Several texts on statistical mechanics
cover these topics in detail; see e.g.,~Ref.~\cite{huang87}.

In finite systems, differences appear between calculations carried 
out using the three ensembles. These differences, dubbed finite-size
effects, have to do with the effect of energy and particle number
fluctuations, and with boundary effects. For example, to describe an
isolated system with mean energy $E$, it is most appropriate to 
use the microcanonical ensemble with that energy. However, 
one can also use a canonical ensemble at a temperature $T$ for which 
the mean energy is $E$. Since the systems used to construct the 
canonical ensemble have different energies from the ones used to 
construct the microcanonical ensemble, one finds differences in the
predictions of each ensemble.  Remarkably, one can show that energy
fluctuations in the canonical ensemble typically scale as the square
root of the volume of the system, whereas the average energy scales 
as the volume of the system.  Hence, the ratio between the energy
fluctuations and the average energy scales as the inverse of the
square root of the volume, and vanishes in the thermodynamic
limit. One then finds that differences between the predictions of 
each ensemble decrease polynomially with increasing volume
(at fixed density). The same applies if one considers the grand
canonical ensemble, where particle number fluctuations typically 
scale with the square root of the volume of the system. Indeed, 
explicit calculations in one dimensional (1D) lattices have shown 
that the differences between the predictions of the canonical and
grand-canonical ensembles for various observables decrease with the
inverse of the number of particles (or lattice sites) in the system
\cite{lebowitz61,rigol_05}.

Experiments usually deal with thermodynamically large systems, whereas
numerical analyses of many-body interacting systems can generally be
done for only (relatively) much smaller system sizes.  Hence, when
trying to theoretically predict/reproduce the outcome an experimental
measurement, a question of much relevance is: \emph{which ensemble should 
one use to minimize finite size effects and obtain the ``thermodynamic 
limit'', or, experimental result?} From the previous discussion about 
the differences between ensembles, one might naively conclude that finite 
size effects always scale polynomially with system size and that, therefore, 
the best one can do theoretically is to optimize exponents and prefactors. 

In this article we show that this is not the case. There is a
preferred ensemble (the grand canonical ensemble) and preferred boundary 
conditions (periodic boundary conditions, so that the system is translationally 
invariant) for which finite-size effects are exponentially small in the 
system size. This holds if the system of interest is in an unordered 
(i.e., without long or quasi-long range order) phase at finite temperature. 
We also consider a different approach to calculating finite-temperature 
properties of many-particle systems, namely, numerical linked cluster 
expansions (NLCEs) \cite{rigol_bryant_06,rigol_bryant_07,rigol_bryant_07b}. 
We show that NLCEs not only exhibit exponential convergence with
increasing system size but generally outperform grand canonical
ensemble calculations in systems with periodic boundary conditions.

The paper is organized as follows. In Sec.~\ref{sec:gener-cons}
we argue, based on a high temperature expansion of the partition
function, that grand canonical ensemble calculations in
translationally invariant systems have exponentially small finite-size
errors. In Sec.~\ref{sec:verification}, we discuss analytically
solvable examples, the 1D and 2D Ising models, that substantiate the
arguments in Sec.~\ref{sec:gener-cons}. In Sec.~\ref{sec:pert-th}, we present 
a proof that finite-size errors are indeed exponentially small in the grand
canonical ensemble for translationally invariant noninteracting systems 
and that, within perturbation theory, the same scaling applies to interacting 
systems. We then study numerically, in Sec.~\ref{sec:numerical-tests}, 
three examples where we systematically compare results from canonical and 
grand canonical ensemble calculations, each for open boundary conditions (OBC) 
and periodic boundary conditions (PBC), and NLCEs. We
summarize our results and conclude in Sec.~\ref{sec:conclusions}.



\section{General considerations}
\label{sec:gener-cons}

In this section, we argue that the GE for a translationally invariant
system (we abbreviate the CE [GE] with open and
  periodic boundary conditions as CE-O [GE-O] and CE-P [GE-P]
  respectively) has exponentially small finite size corrections.  For
that, we make use of a $\beta$ expansion of the free energy, where
$\beta=(k_{B}T)^{-1}$ is the inverse temperature and $k_{B}$ is the
Boltzmann constant. This kind of expansion has been used extensively
in the literature to compute partition functions for various
models \cite{domb60a,domb60b,oitmaa_hamer_book_06}.

Consider the Taylor expansion of the grand partition function $Z\equiv
\Tr e^{-\beta \hat{H}}$ (we set $\mu=0$ for brevity; all the arguments
below are valid for nonzero $\mu$, which will be required for bosons
to prevent Bose-Einstein condensation):
\begin{equation}
  Z(\beta) = \Tr(1) - \beta \Tr(\hat{H}) + \frac{\beta^{2}}{2!}\Tr(\hat{H}^{2}) + \ldots
\end{equation}
We are interested in $\ln Z$, from which thermodynamic quantities can
be obtained by taking suitable $\beta$ or $\mu$ derivatives,
\begin{multline}
  \label{eq:9}
  \ln Z(\beta) = \ln\Tr(1) - \beta \frac{\Tr(\hat{H})}{\Tr(1)} + \\
  + \frac{\beta^{2}}{2}
  \left[\frac{\Tr(\hat{H}^{2})}{\Tr(1)}-\frac{\Tr(\hat{H})^{2}}{\Tr(1)^{2}}\right]
  + \cdots
\end{multline}
We note that
\begin{equation}
  \label{eq:8}
  \frac{\Tr(\hat{H}^{n})}{\Tr(1)} = \frac{\Tr(\hat{H}^{n}e^{-0\cdot \hat{H}})}
  {\Tr(e^{-0\cdot \hat{H}})},
\end{equation}
is an infinite temperature expectation value. At infinite temperature
all unconnected parts of the system, however close to each other, are
uncorrelated. Therefore, the expansion in Eq.~\eqref{eq:9} reduces to
a sum over only the connected graphs that can be embedded in the
finite system \cite{domb60a,sykes66,huang87}. In the CE, the particle
number constraint, i.e., that the total particle number is fixed,
implicitly correlates unconnected pieces of a graph, and therefore
this simplification does not occur.

For a system that has no ordered phase or, equivalently, where order appears 
only at $T=0$, and provided there are no singularities in the complex plane 
\cite{francesco}, the above expansion must converge beyond a certain order. 
If convergence occurs for any temperature $\beta < \infty$, then it must 
come from the coefficients of the $\beta$-expansion, i.e., from the traces 
in Eq.~\eqref{eq:9}.  In other words, for the expansion $(\ln Z)/N = a_{0}+a_{1}\beta 
+ a_{2}\beta^{2}+\cdots$, the coefficient $a_{n}$ must fall faster than any 
exponential for the series to converge for any $\beta$. The convergence cannot 
come from cancellation of terms of opposite sign, since any such cancellation 
can work only at some fine-tuned value of $\beta$.

For a system that has a phase transition between an unordered high temperature 
phase and an ordered low temperature phase, or a singularity in the complex plane, 
preventing convergence beyond an inverse temperature $\beta_{c}$, the coefficients 
do not exhibit this behavior -- in the critical phase the correlation length is 
infinite and all orders of this expansion are relevant. The convergence of the 
series for $\beta<\beta_{c}$ instead comes from the fact that $\beta/\beta_{c} < 1$.  
We verify these arguments in the 1D and 2D Ising models.

From here on, we assume that we are in a regime where the
$\beta$-expansion converges. We will now show that with PBC (when the
system is translationally invariant), all orders of the expansion
Eq.~\eqref{eq:9} up to the system size (to be properly defined below)
are identical to those in the thermodynamic limit. We will further
show that this is not the case with open boundary conditions.

\subsection{Periodic boundary conditions}

Consider a system with $N$ sites and periodic boundary conditions (a
system that is translationally invariant). The $\beta$ expansion of
$\ln Z$ is shown in Eq.~\eqref{eq:9}, in which each term can be
represented by a graph embedded on the finite system. We will call
these graphs clusters \cite{huang87}.  First note that each cluster
has $N$ equivalent positions on the lattice since the system is
translationally invariant.  That gives a factor of $N$ that we move to
the left hand size in Eq.~\eqref{eq:9} to get $(\ln Z)/N$, i.e., an
intensive quantity. Let us now consider a cluster with $c$
sites. First, since we have a cumulant expansion, as discussed above,
only connected clusters enter (see, e.g.,
Refs.~\cite{huang87,domb60a,oitmaa_hamer_book_06} for details).  If
the extent of our cluster in each direction is less than the system
size in that directions (say $L$), then the cluster has open boundary
conditions. Furthermore, even if the system size is increased in any
direction, this cluster is present. Hence, this cluster is present in
the thermodynamic limit. In general, \emph{every cluster with $c$
sites that, in the finite lattice with $N$ sites, does not wrap
around any boundary appears in the infinite system, and vice versa}.
Therefore the contribution of this cluster in a finite system is
exactly the same as its contribution in the thermodynamic limit. On
the other hand, a cluster with $L$ sites in any given direction wraps
around a boundary, i.e., it does not appear in the thermodynamic
limit. As a result, clusters that wrap around boundaries give
contributions that are not present in the thermodynamic limit
\cite{lebowitz61}. Hence, the difference between results in the
thermodynamic limit and in finite-size periodic systems is
$O(\beta^{L-p})$, $p$ being determined by the Hamiltonian.  We note
that $p$ is $O(1)$ for local Hamiltonians, which are the ones of
interest here. Furthermore, based on the earlier argument, we must have
that the coefficient at this order falls faster than any exponential 
in $L-p$ or that the expansion parameter ($\beta/\beta_{c}$) is 
smaller than one. Therefore, finite-size errors in a GE calculation 
of a translationally invariant system at any temperature in the 
unordered phase (where the high-temperature expansion converges) 
are exponential or smaller in $L$, for systems with linear 
dimension $L$.

\subsection{Open boundary conditions}

For a system with OBC, one immediately realizes that clusters do not
have $N$ equivalent positions on the lattice. As a result, even if a
given cluster in the finite system appears in the thermodynamic limit,
its contribution in the finite system will differ from the
thermodynamic limit. For example, for a lattice model in which the
Hamiltonian is a sum of terms involving only nearest neighbor sites,
the term linear in $\beta$ in Eq.~\eqref{eq:9} for a system with OBC
has a correction $O(A/2N)$ relative to the result for PBC, where $A$
is the number of sites in the boundary.  This correction vanishes as
$1/L$ with increasing system size.  Complicated geometric and
combinatorial factors appear at higher orders, all of which approach
the thermodynamic limit result with increasing system size. Hence,
none of the coefficients of a $\beta$ expansion for a finite system
with OBC match the result in the thermodynamic limit, and
finite size errors in $(\ln Z)/N$ are $O(1/L)$.

\subsection{Numerical linked cluster expansions}

Rather than making calculations of finite systems with periodic or
open boundary conditions, and then extrapolating the results to the
thermodynamic limit, another way to calculate finite-temperature
properties of lattice systems in the thermodynamic limit is to use
NLCEs
\cite{rigol_bryant_06,rigol_bryant_07,rigol_bryant_07b,tang_khatami_13}.
The idea in this case is to directly use the linked cluster expansion
of the infinite (translationally invariant) system, for which any
extensive quantity ${\cal O}$ per site can be computed as the sum
\begin{equation}
  \label{eq:nlce1}
  \frac{\cal O}{N} = \sum_{c}M(c)\times W_{\cal O}(c),
\end{equation}
over all connected clusters $c$ that can be embedded in the infinite
lattice. In Eq.~\eqref{eq:nlce1}, $M(c)$ is the \emph{multiplicity} of
cluster $c$, namely, the number of ways per site in which cluster $c$
can be embedded on the lattice, and $W_{\cal O}(c)$ is the
\emph{weight} of the cluster $c$ for observable ${\cal O}$.  
$W_{\cal O}(c)$ is calculated by an inclusion-exclusion principle, one
systematically subtracts contributions from the connected subclusters 
of $c$ \cite{sykes66}
\begin{equation}\label{eq:nlce2}
  W_{\cal O}(c) = \mathcal{O}(c) - \sum_{s\subset c}W_{\cal O}(s).
\end{equation} 
${\cal O}(c)$ is the value of the observable evaluated on the cluster
$c$. In NLCEs, ${\cal O}(c)$ is obtained using a full exact
diagonalization of the Hamiltonian for cluster $c$.

Due to computational limitations, only a finite number of clusters can
ultimately be calculated in Eq.~\eqref{eq:nlce1}. Nevertheless, as
shown in
Refs.~\cite{rigol_bryant_06,rigol_bryant_07,rigol_bryant_07b}, NLCEs
can converge at lower temperatures than high temperature expansions,
and sometimes all the way to the ground state for systems with
unordered ground states. Also, NLCEs can provide very accurate
results for temperatures at which exact diagonalization results for
systems with periodic boundary conditions suffer from very large
finite-size effects. A pedagogical introduction to implementing NLCEs
can be found in Ref.~\cite{tang_khatami_13}.

In what follows, we compare NLCE results with those obtained in
calculations in finite systems with different boundary conditions.
Our goal is to find how each of them converges to the thermodynamic
limit result and which converges the fastest. For NLCEs, the accuracy
of the results is determined by the size of the largest clusters
considered in the sum in Eq.~\eqref{eq:nlce1} and the model under
consideration.

\section{Verification in Ising models}
\label{sec:verification}

In this section, we verify the arguments given in
Sec.~\ref{sec:gener-cons} in the 1D and 2D Ising models, both of which
can be solved analytically.

\subsection{1D Ising model}
\label{sec:1d-ising}

In the thermodynamic limit, the log of the partition function per
site, $\Omega$, can be obtained using the transfer matrix method
\cite{huang87}, and is given by (we set $J=1$)
\begin{equation}
  \label{eq:1}
  \Omega(\beta) = \ln(e^{\beta} + e^{-\beta}).
\end{equation}
The expansion in powers of $\beta$ of this result is
\begin{equation}
  \label{eq:4}
  \Omega(\beta) = \ln2 + \frac{\beta^{2}}{2}-\frac{\beta^{4}}{12} 
  + \frac{\beta^{6}}{45}-\frac{17\beta^{8}}{2520}
  + \frac{31\beta^{10}}{14175}+ \cdots
\end{equation}

The result for finite systems with periodic boundary conditions is
given by
\begin{equation}
  \label{eq:2}
  \begin{split}
    \Omega_L(\beta) &= \frac{1}{L}\ln \left[(e^{\beta} +
      e^{-\beta})^{L}
      + (e^{\beta} - e^{-\beta})^{L}\right] \\
    &= \Omega(\beta) + \frac{\tanh^{L}\beta}{L} + \cdots
  \end{split}
\end{equation}
Since $0\leq \tanh\beta<1$ for $0\leq \beta<\infty$, the finite size
error is $1/L$ times an exponentially small number in $L$. Quantities 
like the energy, which are derivatives of the free energy, converge 
exponentially fast with $L$.

Expanding Eq.~\eqref{eq:2} in powers of $\beta$ for different values
of $L$, we get
\begin{equation}
  \label{eq:3}
  \begin{split}
    \Omega_2(\beta) &= \ln2 + \beta^{2} + \ldots \\
    \Omega_3(\beta) &= \ln2 + \frac{\beta^{2}}{2} + \frac{\beta^{3}}{3} + \ldots \\
    \Omega_4(\beta) &= \ln2 + \frac{\beta^{2}}{2} + \frac{\beta^{4}}{6} + \ldots \\
    \Omega_5(\beta) &= \ln2 + \frac{\beta^{2}}{2} -
    \frac{\beta^{4}}{12} +
    \frac{\beta^{5}}{5} + \ldots \\
    \Omega_6(\beta) &= \ln2 + \frac{\beta^{2}}{2} -
    \frac{\beta^{4}}{12} + \frac{17\beta^{6}}{90} + \ldots
  \end{split}
\end{equation}
As one can see, the results are exact to $O(\beta^{L-1})$. Naturally,
as one goes to lower temperatures, the correlation length increases
and larger systems are required to capture the relevant powers of
$\beta$.  It is easy to verify that, with open boundary conditions,
the corrections are always $O(1/L)$.

\begin{figure}[!tb]
  \centering
  \includegraphics[width=0.95\columnwidth]{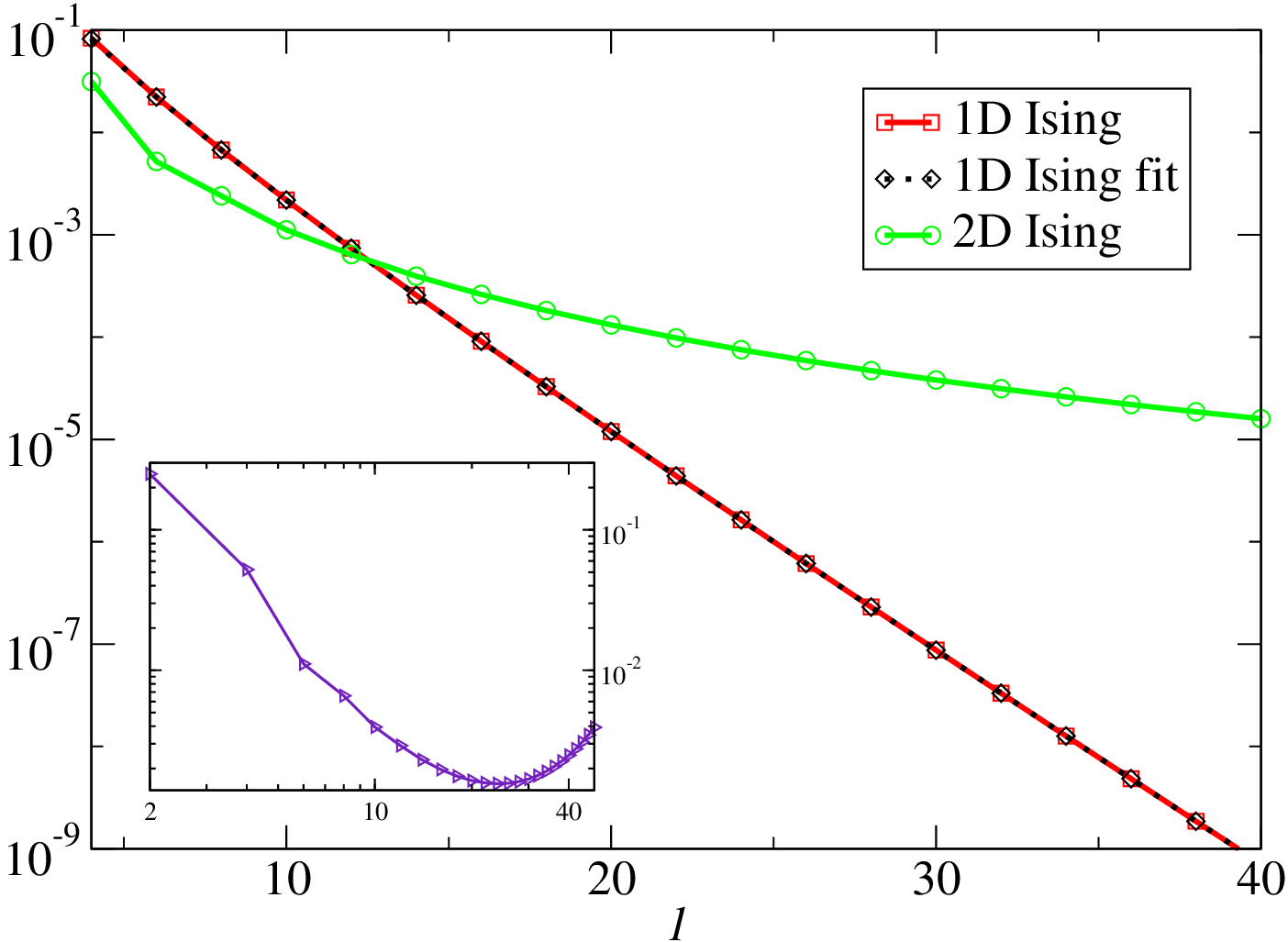}
  \caption{(Color online) Coefficients (absolute value) of the
    $\beta$-expansion of the free energy vs. the expansion order
    $l$. We show the coefficients in Eq.~\eqref{eq:4} for the 1D Ising
    model, and a fit to the function $a b^{-l}/l$ with $a=2.0000$ and
    $b=1.5708$ (the difference between the exact result and
    $2(\pi/2)^{-l}/l$ vanishes exponentially fast with $l$), and from
    Eq.~\eqref{eq:7} for the 2D Ising model. The coefficients in the
    latter case do not fall off exponentially fast. Because of this,
    the $\beta$-expansion only converges for $\beta < \beta_{c}$. It
    is only in this regime that finite-size errors in grand-canonical
    ensemble calculations of translationally invariant systems are
    exponentially small in system size. \textbf{Inset:}
      Shows the rational part of the coefficients in
      Eq.~\eqref{eq:7}. They decrease at first but increase after
      $O(\eta^{24})$. See text for further discussion.}
  \label{fig:ising-fit}
\end{figure}

Figure \ref{fig:ising-fit} shows a plot of the coefficients of
expansion in Eq.~\eqref{eq:4} along with a fit to $a b^{-L}/L$. The fit 
shows that $b=\pi/2$, as expected from the singularity of the partition 
function $\Omega(\beta)$ at $\beta=i\pi/2$.

To conclude our discussion of the 1D Ising model, we evaluate the
first few orders of the NLCE for this model (instead of numerical
exact diagonalization of the clusters, we obtain these results
analytically).  First, we evaluate the partition function [$\ln
Z_L(\beta)$] on finite clusters with OBC
\begin{equation}
  \label{eq:ising-nlce-prop}
  \begin{split}
    \ln Z_1(\beta) &= \ln2, \\
    \ln Z_2(\beta) &= \ln2 + \ln\left(e^{\beta} + e^{-\beta}\right),\\
    \ln Z_3(\beta) &= \ln2 + \ln\left(e^{2\beta}+e^{-2\beta} +
      2\right),
  \end{split}
\end{equation}
and then carry out the subtractions. The weights are given by [see
Eq.~\eqref{eq:nlce2}]
\begin{equation}
  \label{eq:ising-nlce-weights}
  \begin{split}
    W_{1} &= \ln Z_{1}(\beta)  = \ln2, \\
    W_{2} &= \ln Z_{2}(\beta)-2W_{1} = \ln\left(e^{\beta} + e^{-\beta}\right)- \ln2,\\
    W_{3} &= \ln Z_{3}(\beta)-2W_{2}-3W_{1}= 0.
  \end{split}
\end{equation}
Hence, the result for $\Omega$ obtained in calculations including up
to $n$ sites, $(\Omega)_n$, is given by [see Eq.~\eqref{eq:nlce1}]
\begin{equation}
  \label{eq:ising-nlce-sums}
  \begin{split}
    (\Omega)_1 &= W_{1}=\ln2, \\
    (\Omega)_2 &= W_{1}+W_{2}= \ln\left(e^{\beta} + e^{-\beta}\right), \\
    (\Omega)_3 &= W_{1}+W_{2}+W_{3} = \ln\left(e^{\beta} +
      e^{-\beta}\right).
  \end{split}
\end{equation}
It can be verified that the last result is valid at all higher orders
in the ``NLCE''.  The thermodynamic limit result is therefore obtained
by just considering clusters with one and two sites. This is an
infinite improvement over the use of the grand canonical ensemble with
periodic boundary conditions.  Whereas this infinite gain is specific
to the 1D Ising model --- the model can, after all, be solved using a
two dimensional transfer matrix, we show in what follows that the fact
that NLCEs outperform exact calculations in finite systems appears to
be generic.

\subsection{2D Ising model}
\label{sec:2d-ising}

For the 2D Ising model, $\Omega$ is given by \cite{huang87,Onsager_44}
\begin{multline}
  \label{eq:5}
  \Omega(\beta) = \ln[2\cosh(2\beta)] + \\ +
  \int_{0}^{\pi}\frac{d\phi}{2\pi}
  \ln\left[\frac{1+\sqrt{1-\frac{4\sin^{2}\phi}{\cosh^{2}(2\beta)\coth^{2}(2\beta)}}}{2}
  \right].
\end{multline}
The $\beta$-expansion of this result is given by
\begin{equation}
  \label{eq:6}
  \Omega(\beta) = \ln2 + \beta^{2} + \frac{5\beta^{4}}{6} + \frac{32\beta^{6}}{45} 
  + \frac{425\beta^{8}}{252} + \ldots
\end{equation}
One can see that the coefficients become larger than 1 for higher
orders. The correct expansion parameter for models with a
finite-temperature transition is $\beta/\beta_{c}$.  For the classical
2D Ising model on a square lattice, $\beta_{c} = \ln(1+\sqrt{2})/2$.
This gives (with $\eta \equiv \beta/\beta_{c}$),
\begin{equation}
  \label{eq:7}
  \Omega(\beta) = \ln2 + \frac{a^{2}\eta^{2}}{4} + \frac{5a^{4}\eta^{4}}{96} + 
  \frac{a^{6}\eta^{6}}{90} + \frac{425a^{8}\eta^{8}}{64512} + \ldots
\end{equation}
where $a\equiv\ln(1+\sqrt{2})<1$. The coefficients of the $\eta$-expansion 
are plotted versus the order of the expansion in Fig.~\ref{fig:ising-fit}.  
They do not fall off faster than exponential, or exponentially. 
We note that the rational part of the coefficients of the 
first few orders of the $\eta$-expansion reported in Eq.~\eqref{eq:7} is 
deceiving. They decrease with increasing order of  
expansion. This, together with the fact that $a<1$, suggests that the 
coefficients of the $\eta$-expansion should fall off 
faster than exponentially. However, as shown in the inset in 
Fig.~\ref{fig:ising-fit}, the aforementioned rational part \emph{increases} 
with increasing order of expansion after $O(\eta^{24})$. Because of this,
convergence in the $\eta$ expansion is only expected for $\eta<1$ and 
does not come from the coefficients.

We also calculate $\ln Z_{L\times L}$ for small systems with $N=L\times
L$ sites in the GE for both periodic and open boundary
conditions. With PBCs
\begin{eqnarray}
  \label{eq:2d-ising-g-p}
  \ln Z^{\rm 2D-P }_{1\times1} &=& \ln2, \nonumber\\
  \ln Z^{\rm 2D-P}_{2\times 2} &=& \ln2 + \ln\left(e^{4\beta} + e^{-4\beta}+2\right), \\
  \ln Z^{\rm 2D-P}_{3\times 3} &=& \ln2 + \ln\bigl(e^{-18 \beta
  }+9 e^{-10 \beta }+24 e^{-6 \beta }+ \nonumber\\
  && +99 e^{-2 \beta }+72 e^{2
    \beta }+51 e^{6 \beta }\bigr),\nonumber
\end{eqnarray}
whereas, with OBCs,
\begin{eqnarray}
  \label{eq:2d-ising-g-o}
  \ln Z^{\rm 2D-O}_{1\times1} &=& \ln2 \nonumber\\
  \ln Z^{\rm 2D-O}_{2\times 2} &=& \ln2 + \ln\left(e^{4\beta} + e^{-4\beta}+2\right) \\
  \ln Z^{\rm 2D-O}_{3\times 3} &=& \ln2 + \ln\bigl(e^{-12 \beta
  }+4 e^{-8 \beta }+16 e^{-6 \beta }+\nonumber\\ 
  &&+23 e^{-4 \beta }+48 e^{-2
    \beta }+
  48 e^{2 \beta }+23 e^{4 \beta }+\nonumber\\
  &&+ 16 e^{6 \beta }+4 e^{8
    \beta }+ e^{12 \beta }+72 \bigr)\nonumber.
\end{eqnarray}

For the $\beta$-expansion of the $3\times 3$ systems, up to the first
term that differs from Eq.~\eqref{eq:6}, we obtain
\begin{equation}
  \label{eq:2d-ising-finite-expansion}
  \begin{split}
    \Omega^{\rm 2D-O}_{3\times 3} &= \ln2 + \frac{2\beta^{2}}{3} +
    \ldots \\
    \Omega^{\rm 2D-P}_{3\times 3} &= \ln2 + \beta^{2} +
    \frac{2\beta^{3}}{3} + \ldots
  \end{split}
\end{equation}
We see that whereas for OBC the coefficient of the second order term is
incorrect, for PBC it is correct, i.e., once again GE-P gives results
that are correct to $O(\beta^{L-1})$, with $L=3$ in this case.

For the NLCE calculation with clusters with up to four sites, we
obtain
\begin{eqnarray}
  \label{eq:2d-ising-nlce}
  &&(\Omega)_4 = 20 \ln \left(e^{-\beta }+e^{\beta }\right)+  54
  \ln \left(e^{- \beta } e^{2 \beta }+1\right)\nonumber\\
  && \ -38
  \ln \left(e^{-2 \beta }+e^{2 \beta }+2\right)+\ln \left(e^{-4
      \beta }+e^{4 \beta
    }+6\right).\
\end{eqnarray}
Expanding in powers of $\beta$, and reporting terms up to the first
one that differs from Eq.~\eqref{eq:6}, we get
\begin{equation}\label{eq:2dinlcee}
  (\Omega)_4 = \ln2 + \beta^{2} + \frac{5\beta^{4}}{6} -
  \frac{58\beta^{6}}{45} + \ldots
\end{equation}
The above result is correct to $O(\beta^{5})$. We must stress that
Eq.~\eqref{eq:2dinlcee} was obtained in an expansion in which the
largest cluster has $N=4$, while
Eqs.~\eqref{eq:2d-ising-finite-expansion} are for systems with
$N=9$. The gain is evident.

\section{Finite temperature perturbation theory to all orders}
\label{sec:pert-th}

In the case of bosons or fermions on a lattice that can be treated by
finite-temperature perturbation theory (with a noninteracting theory
as the unperturbed starting point), a proof that finite-size errors are 
exponentially small to all orders in perturbation theory can be made based 
on the momentum-space representation of the Hamiltonian. The proof is 
essentially identical for bosons and fermions, so we focus on the former.

We consider a generic massive scalar field theory in a 1D lattice,
with unit lattice spacing, $L$ sites, and PBC:
\begin{equation}
  \label{H}
  \hat{H}_{s}=\frac12 \sum_{j=1}^L\left[\hat{\pi}_j^2
    +(\hat{\ph}_{j+1}-\hat{\ph}_j)^2 + m^2\hat{\ph}_j^2\right],
\end{equation}
where $[\hat{\ph}_j,\hat{\pi}_{j'}]=i\delta_{jj'}$ and
$\hat{\ph}_{j+L}\equiv\hat{\ph}_j$,
$\hat{\pi}_{j+L}\equiv\hat{\pi}_j$.  
This Hamiltonian is diagonalized via
\begin{equation}
  \begin{split}
    \hat{\ph}_j &= \frac{1}{\sqrt{2L}}\sum_{n=0}^{L-1}\w_n^{1/2}
    \left[e^{\frac{2\pi i n j}{L}}\hat{a}_n + e^{-\frac{2\pi i n
          j}{L}}\ad_n\right],\\
    \hat{\pi}_j &= -\frac{i}{\sqrt{2L}}\sum_{n=0}^{L-1}\w_n^{-1/2}
    \left[e^{\frac{2\pi i n j}{L}}\hat{a}_n - e^{-\frac{2\pi i n j}{L}}\ad_n\right],
  \end{split}
\end{equation}
so that
\begin{equation} \label{Hk} \hat{H}_{s} = \sum_{n=0}^{L-1}\w_n\ad_n
  \hat{a}_n + \text{constant}\hfill,
\end{equation}
where $[\hat{a}^{}_n,\ad_{n'}] = \delta_{nn'}$, $\w_n =\w(k_n)$, $k_n = 2\pi n/L$, 
$n\in[0,L-1]$, and
\begin{equation}
  \label{wk}
  \w(k) = \sqrt{2(1-\cos k)+m^2} =
  \sqrt{4\sin^2(k/2)+m^2}.
\end{equation}

We now want to compute the grand canonical partition function
\begin{equation}
  Z(\beta,\mu)\equiv \mathop{\rm Tr}e^{-\beta (\hat{H} -\mu \hat{N})}\,,
  \label{Z}
\end{equation} 
where $\hat{N}=\sum_n \ad_n \hat{a}_n$ is the total number operator.
We take the trace in the Fock basis of eigenstates of each $\ad_n a_n$,
\begin{equation}
  Z_L(\beta,\mu) = \prod_{n=0}^{L-1}\sum_{N_n=0}^\infty
  e^{-\beta(\w_n-\mu)N_n} = \prod_{n=0}^{L-1} \frac{1}{1-e^{-\beta(\w_n-\mu)}}.
  \label{Z1}
\end{equation} 
Equivalently,
\begin{equation}
  \Omega_L(\beta,\mu) = \frac{1}{L}\ln Z_L(\beta,\mu) = 
  \frac{1}{L}\sum_{n=0}^{L-1} F(k_n),
  \label{OL}
\end{equation} 
where we have defined
\begin{equation}
  \label{F}
  F(k) \equiv -\ln[1-e^{-\beta(\w(k)-\mu)}].
\end{equation}
In the thermodynamic limit, the sum over $n$ becomes an integral,
\begin{equation}
  \Omega(\beta,\mu)\equiv\lim_{L\to\infty}\Omega_L(\beta,\mu)= 
  \frac{1}{2\pi}\int_{0}^{2\pi}dk\,F(k)\,.
  \label{Oi}
\end{equation}

We now wish to show that $|\Omega_L(\beta,\mu)-\Omega(\beta,\mu)|$ is
exponentially small in $L$.

We first note that $F(k)$ is periodic in $k$ with period $2\pi$.
Therefore its Fourier expansion takes the form
\begin{equation}
  F(k)=\sum_{j=-\infty}^{+\infty}\tilde F_j\,e^{i jk}\,,
  \label{fsum}
\end{equation} 
where the Fourier coefficients are given by
\begin{equation}
  \tilde F_j = \frac{1}{2\pi} \int_{0}^{2\pi}dk\,e^{-i jk}F(k)\,.
  \label{tildef}
\end{equation} 
Using \eqs{Oi} and (\ref{tildef}), we get
\begin{equation}
  \Omega(\beta,\mu) = \tilde F_0\,.
  \label{Oi1}
\end{equation} 
Using \eqs{OL} and (\ref{fsum}), we get
\begin{eqnarray}
  \Omega_L(\beta,\mu) &=& \frac{1}{L}\sum_{n=0}^{L-1} F(2\pi n/L)
  = \frac{1}{L}\sum_{n=0}^{L-1}\sum_{j=-\infty}^{+\infty}\tilde
  F_j\,e^{2\pi i jn/L} \nonumber\\
  &=& \sum_{j=-\infty}^{+\infty}\tilde F_j\biggl[\frac{1}{L}
  \sum_{n=0}^{L-1}e^{2\pi i jn/L}\biggr]  \label{OL5}
  \\& =& \sum_{j=-\infty}^{+\infty}\tilde
  F_j\delta_{j\,\rm{mod}\,L,\,0}
  = \sum_{j'=-\infty}^{+\infty}\tilde F_{j'\!L}\,.\nonumber
\end{eqnarray}
Subtracting \eq{Oi1} from \eq{OL5}, we get \begin{equation}
  O_L(\beta,\mu)-\Omega(\beta,\mu) = \sum_{j\ne 0}\tilde F_{jL}\,.
  \label{Odiff}
\end{equation} 
Examining \eqs{wk} and (\ref{F}), we see that if $m>0$ and $\mu<m$
(necessary to avoid Bose condensation), then $F(k)$ is continuous and
infinitely differentiable for all real $k$.  It then follows from a
general theorem of Fourier series \cite{tolstov76} that $\tilde F_{jL}$ goes
to zero faster than any power of $|j|L$ as $L\to\infty$.  The sum over
$j$ in \eq{Odiff} will then be dominated by the $j=\pm1$ terms.  We
conclude that $|O_L(\beta,\mu)-\Omega(\beta,\mu)|$ is exponentially
small in $L$ if $m>0$ and $\mu<m$.

We can verify this explicitly. Again examining \eqs{wk} and (\ref{F}),
we see that $F(k)$ is changing most rapidly near $k=0$. For $m>0$ and
small enough $k$, $F(k)$ can be approximated via
\begin{equation}
  F(k)\simeq -\ln[1-e^{-\beta(m-\mu)}e^{-\beta k^2/2m}].
  \label{F0}
\end{equation} 
Assuming $\mu<m$, we have
\begin{equation}
  F(k)\simeq \sum_{n=1}^\infty\frac{1}{n}e^{-n\beta(m-\mu)}e^{-n \beta k^2/2m}\,.
  \label{F2}
\end{equation} 
In this approximation, we get
\begin{equation}
  \tilde F_{jL} \simeq
  \sqrt{\frac{m}{2\pi\beta}} \sum_{n=1}^\infty \frac{1}{n^{3/2}}
  e^{-n\beta(m-\mu)}e^{-j^2 L^2 m/2 \beta n}\,.
  \label{tF2}
\end{equation} 
The sum over $n$ can be approximated by steepest descent.  For
$j^2L^2\gg1/m(m-\mu)$, we find
\begin{equation}
  \tilde F_{jL} \simeq  \frac{1}{|j|L} \exp\Bigl[-[2m(m-\mu)]^{1/2}|j|L\Bigr]\,,
  \label{tF3}
\end{equation} 
This is exponentially small in $L$, as expected from the general
theorem.  The sum over $j$ in \eq{Odiff} is then dominated by
$j=\pm1$, and we have
\begin{equation}
  |\Omega_L(\beta,\mu)-\Omega(\beta,\mu)|
  \simeq \frac{2}{L}\exp\Bigl[-[2m(m-\mu)]^{1/2}L\Bigr]\,,
  \label{Odiff2}
\end{equation}
which is the behavior observed for the 1D Ising model in
Fig.~\ref{fig:ising-fit}.

This proof extends straightforwardly to higher dimensions, assuming
periodic boundary conditions in each dimension.

Now consider adding an interaction term to $\hat{H_s}$, such as
$g\sum_j\ph_j^4$.  We can compute $\Omega_L(\beta,\mu)$ order by order
in finite temperature perturbation theory.  Each term is represented by 
a connected Feynman diagram \cite{fetter_bk_03}. The expression for a diagram with $p$ 
propagators and $v$ vertices takes the form 
\begin{equation} \frac{1}{L}\sum_{n_1=0}^{L-1}
  \ldots \frac{1}{L}\sum_{n_{p-v+1}=0}^{L-1} g^v
  F(k_{n_1},\ldots,k_{n_{p-v+1}})\,.
  \label{feyn}
\end{equation} 
For $m>0$ and $\mu<m$, $F$ is infinitely differentiable in each
$k_{n_i}$. Hence, one can once again apply the general theorem that says
that the difference between \eq{feyn} and the $L\to\infty$ limit must
be exponentially small in $L$.

Other models of bosons and/or fermions can be analyzed in exactly the same
way, with only the form of $F(k)$ changing. As long as $F(k)$ is infinitely 
differentiable, the argument holds.  

\section{Numerical tests}
\label{sec:numerical-tests}

In this section, we discuss numerical results that support the
generality of the conclusions reached so far. The results are obtained
using full exact diagonalization calculations for fermionic systems in
one dimension. We study spinless fermions with nearest-neighbor
hoppings that are either noninteracting or that have nearest and
next-nearest-neighbor interactions. They are described by the
following generic Hamiltonian
\begin{multline}
  \label{eq:gen_f_ham}
  \hat{H}_{\rm f} =
  \sum_{i}\Biggl[-t(\hat{c}^{\dag}_{i}\hat{c}^{}_{i+1}+\text{H.c.})
  + V \left(\hat{n}_{i}-\frac12\right)\left(\hat{n}_{i+1}-\frac12\right) \\
  + V'\left(\hat{n}_{i}-\frac12\right)\left(\hat{n}_{i+2}-\frac12\right)\Biggl],
  \qquad\quad
\end{multline}
where $\hat{n}^{}_{i}=\hat{c}^{\dag}_{i}\hat{c}^{}_{i}$ is the number
operator.  We present results for different values of the parameters
$V$ and $V'$, and using both the canonical and grand canonical
ensembles with open and periodic boundary conditions. We also report
NLCE results.

\begin{figure}[!t]
  \centering
  \includegraphics[width=0.98\columnwidth]{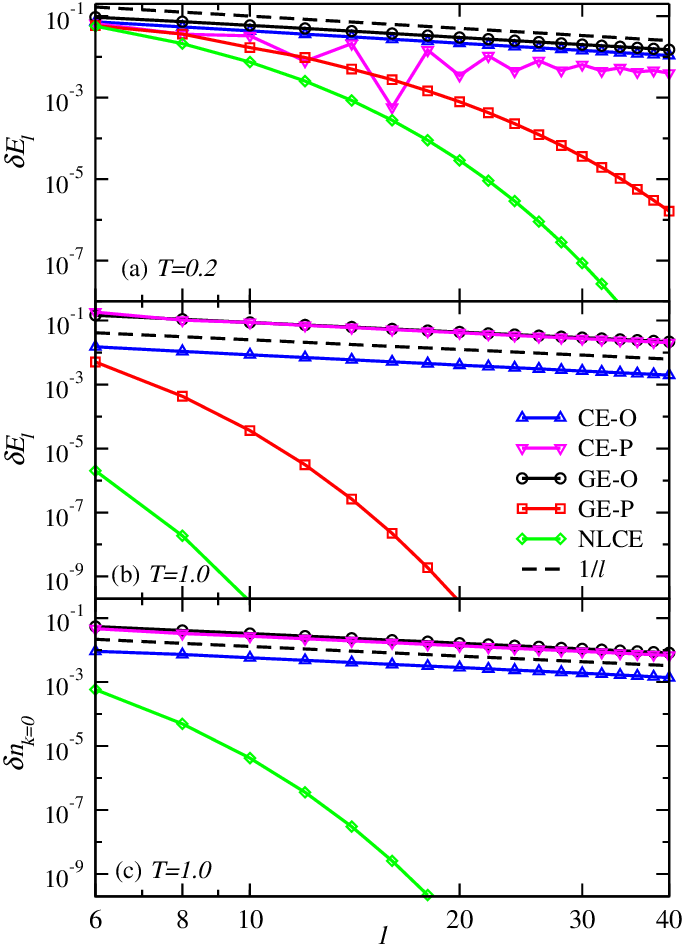}
  \caption{(Color online) Comparison of results of canonical and grand canonical
    ensembles with open and periodic boundary conditions, and NLCEs,
    for noninteracting fermions hopping on a one dimensional lattice
    (corresponds to Eq.~\eqref{eq:gen_f_ham} with $t=1$,
    $V=V'=0$). The energy difference $\delta E_l$ (see text) is
    plotted for two different temperatures in (a) and (b). Panel (c)
    shows the occupation of the $k=0$ mode, or, equivalently, the
    correlation $L^{-1}\sum_{ij}\langle c^{\dag}_{i}c_{j}\rangle$.
    $\delta n_{k=0}$ is zero for the grand-canonical ensemble with
    PBCs, see text.}
  \label{fig:free-ferm}
\end{figure}

In all cases, we calculate the energy per site $E_{l}$. For
noninteracting fermions, we also compute the occupation of the
momentum $k=0$ mode, $n^l_{k=0}\equiv \sum_{i,j}\langle
\hat{c}^{\dag}_{i}\hat{c}^{}_{j}\rangle$.  For the interacting models,
we compute the the nearest-neighbor single-particle correlation
function $K_l = \sum_{i}\langle
\hat{c}^{\dag}_{i}\hat{c}^{}_{i+1}+{\rm H.c.}\rangle$. For the results
reported, by $l$ we mean the number of sites of the finite system or
the number of sites of the largest cluster in the NLCE. We plot the
finite size errors $\delta E_{l}$, $\delta n^l_{k=0}$, and $\delta
K_{l}$, with $\delta O_{l}\equiv (O_{l}-O)/O$ where $O$
($E,\,n_{k=0},\,K$) is either the exact analytic result in the
thermodynamic limit, when known, or the highest order result from a
numerical linked cluster expansion.

\begin{figure}[!t]
  \centering
  \includegraphics[width=0.98\columnwidth]{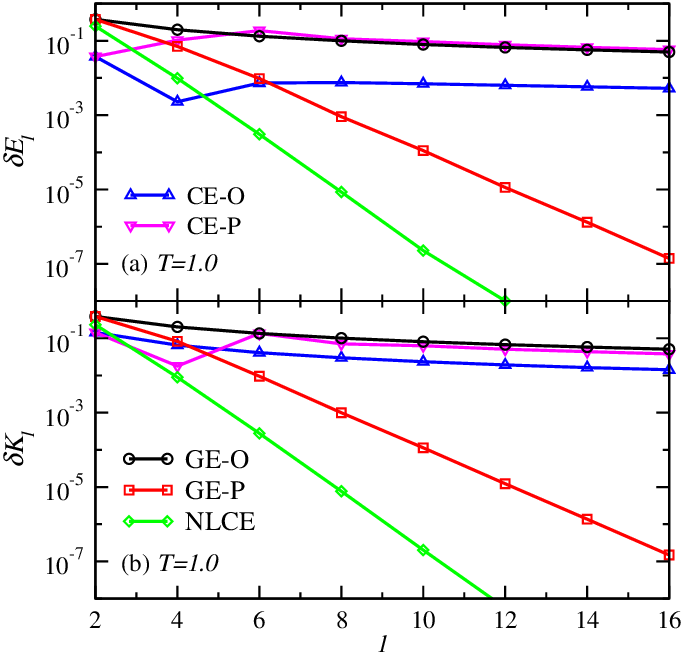}
  \caption{(Color online) (a) $\delta E_{l}$ and (b) $\delta K_{l}$ for interacting
    fermions [Eq.~\eqref{eq:gen_f_ham}] with $t=1$ (unit of energy),
    $V=1.0,\,V'=0$, and $T=1.0$.}
  \label{fig:t1v1}
\end{figure}
\begin{figure}[!b]
  \centering
  \includegraphics[width=0.98\columnwidth]{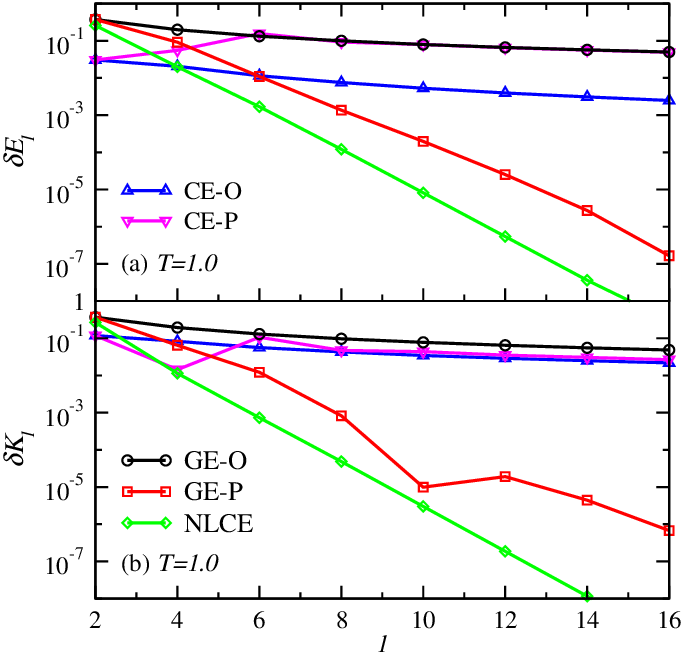}
  \caption{(Color online) (a) $\delta E_{l}$ and (b) $\delta K_{l}$ for interacting
    fermions [Eq.~\eqref{eq:gen_f_ham}] with $t=1$ (unit of energy),
    $V=1.0,\,V'=1.0$, and $T=1.0$.}
  \label{fig:t1v1vp1}
\end{figure}

Figure \ref{fig:free-ferm} reports results for $\delta E_{l}$ and
$\delta n^l_{k=0}$ for noninteracting fermions
[Eq.~\eqref{eq:gen_f_ham} with $t=1$ and $V=V'=0$].  Panels (a) and
(b) show the finite-size errors of the energy at two temperatures, and
(c) shows the finite-size errors of the zero-mode occupation (the sum
of all one-particle correlations). The results for the grand-canonical
energy in finite systems can be obtained analytically
\begin{equation}
  \label{eq:10}
  E_{l} = \sum_{n=0}^{l-1} \frac{\epsilon_{n} e^{-\beta\epsilon_{n}}}
  {1+e^{-\beta\epsilon_{n}}},
\end{equation}
where $\epsilon_{n}$ are the single-particle eigenenergies.  The
momentum distribution function with periodic boundary conditions is
given by the Fermi-Dirac distribution
\begin{equation}
  \label{eq:11}
  n_{k} = \frac{1}{1+e^{-\beta\epsilon_{k}}}.
\end{equation}
We note that $n_{k}$ for finite systems (for the values of $k$ allowed)
is exactly the same as in the thermodynamic limit. Hence, $\delta
n^l_{k=0}=0$ for the grand canonical ensemble with PBC.  Therefore,
no error is reported for the GE-P in Fig.~\ref{fig:free-ferm}(c).  The results in
the canonical ensemble are obtained as described in
Ref.~\cite{rigol_05}.

For the two temperatures and two quantities reported in
Fig.~\ref{fig:free-ferm}, one can see that the errors of the GE-O,
CE-P, and CE-O results decrease as $1/l$. This is expected and is made
apparent in the plots by comparing those results to the $1/l$ plots
(dashed lines) depicted for reference. On the other hand, the GE-P
(for the energy) and NLCE errors can be seen to decrease exponentially
with increasing $l$. For the energy, the NLCE errors are much smaller
than the GE-P ones, showing once again that NLCEs generally ($n_{k}$
for noninteracting fermions being a counterexample) outperform
grand canonical calculations in finite translationally invariant
systems.

\begin{figure}[!t]
  \centering
  \includegraphics[width=0.96\columnwidth]{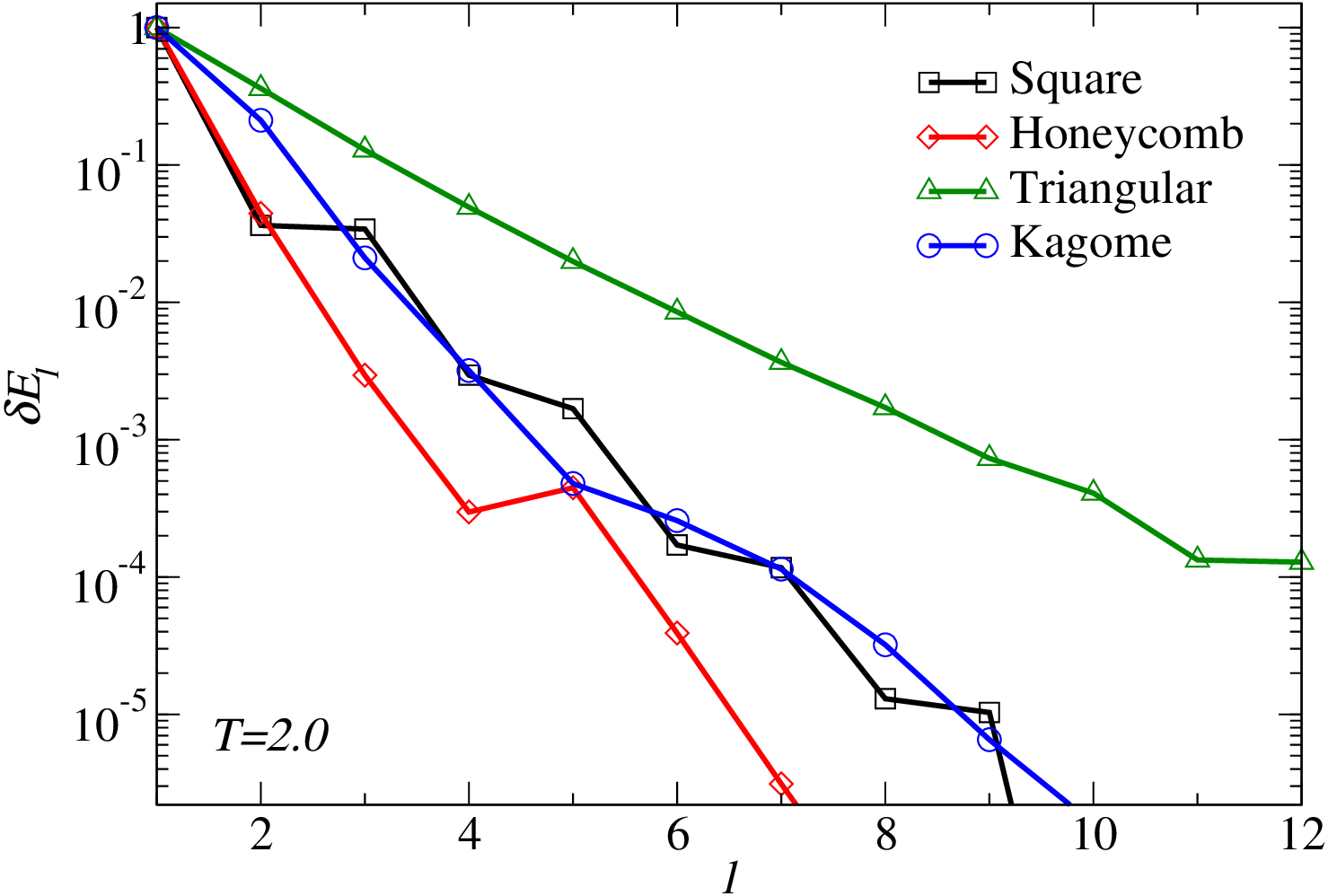}
  \caption{(Color online) $\delta E_{l}$ in NLCE calculations of the Heisenberg model
    in four different two-dimensional lattices, as indicated. The
    approach to the thermodynamic limit result is again exponentially
    fast as one increases the order of the expansion. The results presented 
    here were taken from 
    Refs.~\cite{rigol_bryant_07,tang_khatami_13,tang_paiva_12,tang_paiva_13}.
    $J=1$ is taken to be the unit of energy.}
  \label{fig:2dmodels}
\end{figure}

Figures \ref{fig:t1v1} and \ref{fig:t1v1vp1} show the results for: (a)
$\delta E_{l}$ and (b) $\delta K_{l}$ for interacting fermions
[Eq.~\eqref{eq:gen_f_ham}] with $(t,V,V')=(1,1,0)$ and $(1,1,1)$,
respectively.  In all cases, it is apparent that only the GE-P and
NLCE errors decrease exponentially fast with $l$, while the GE-O and
CE errors decrease as $1/l$. Once again, these results (now for
interacting systems) show that the NLCE errors are significantly
smaller than those of the GE-P. Here, we have used the highest order
of the NLCE ($l=18$) as the estimate for the 
thermodynamic limit \cite{rigol_14}. All results for the interacting
models were obtained using full exact diagonalization of the
Hamiltonian.

Finally, although we cannot make an analysis in 2D equivalent to that
presented for interacting models in 1D (because of the exponential
scaling of the computational cost combined with the fast increase of $N$), 
we can still verify that NLCE calculations
have exponentially small errors. In Fig.~\ref{fig:2dmodels}, we show
results for $\delta E_{l}$ for the 2D Heisenberg model in four
different lattice geometries --- square \cite{tang_khatami_13}, 
honeycomb \cite{tang_paiva_12,tang_paiva_13}, kagome, and triangular
\cite{rigol_bryant_07}. In all cases, the error is once again seen to
decrease exponentially fast with increasing the number of sites in the
clusters considered.

\section{Conclusions}
\label{sec:conclusions}

We have shown that grand canonical ensemble calculations in
translationally invariant systems that are in unordered phases at
finite temperatures have exponentially small finite size errors, whereas
canonical ensemble calculations have errors that are power law in system 
size. Hence, while full exact diagonalization calculations in the
canonical ensemble are computationally less demanding than grand canonical 
ones, if one is interested in accurately computing quantities in an 
unordered finite-temperature phase, using the grand canonical ensemble is
preferable. The additional computational cost incurred by
diagonalizing all particle sectors (around a factor two in fermionic 
systems at half filling) for the GE is far less than the speed up
gained by having to study smaller systems (exponential). Furthermore, we have shown
that numerical linked cluster expansions generally have even smaller
errors than grand canonical ensemble calculations in translationally
invariant systems. The additional cost incurred in the diagonalization
of many clusters with a given size is far less than the speed up
gained by having to study clusters that are much smaller than finite
systems with periodic boundary conditions. The benefit of using NLCEs
is most striking in two-dimensional lattices. 

The specific system sizes (cluster sizes) 
required to observe exponential convergence in grand canonical calculations 
of systems with periodic boundary conditions (in numerical linked cluster expansions) 
depend on details such as the model under consideration and the temperatures 
of interest, which set the relevant correlation length. Otherwise, our 
conclusions are completely general.

\break

\section*{Acknowledgments}
We thank Anders Sandvik and Cristian Batista for stimulating
discussions. This work was supported by the Office of 
Naval Research (D.I. and M.R.), and by NSF Grant PHY13-16748 (M.S.).

\bibliography{FiniteSize}

\maketitle

\end{document}